\documentclass[twocolumn,aps,prd,showpacs,preprintnumbers,nofootinbib]{revtex4}
\usepackage{amsfonts,amssymb,amsmath}
\usepackage{graphicx}

\begin{document}
\preprint{YITP-13-38, RESCEU-10/13}

\title{  
Detecting cosmic string passage through the Earth by consequent global earthquake
}

\author{Hayato~Motohashi$^{1,2}$}  
\author{Teruaki Suyama$^{2}$}  
\address{    
$^{1}$ Yukawa Institute for Theoretical Physics,  
Kyoto University, Kyoto 606-8502, Japan \\
$^{2}$ Research Center for the Early Universe (RESCEU),  
Graduate School of Science, The University of Tokyo, Tokyo 113-0033, Japan 
}  

\begin{abstract}
Effects invoked by the passage of the cosmic string 
through the Earth are investigated.
The cosmic string induces global oscillations of the Earth whose amplitude and
acceleration both linearly depend on the string line density.
For the line density maximally allowed by the cosmic observations,
the oscillations are perceivable even to human beings and may cause serious
damages to the environment.  
Use of the sophisticated accelerograph enables us to detect the string
of a line density down to ten orders of magnitude smaller than the cosmologically
relevant value.
\end{abstract}


\maketitle  

\section{Introduction}
Cosmic strings are line-like topological defects extending over the cosmological scales 
and are ubiquitous in 
field theories with spontaneous symmetry breaking~\cite{Jeannerot:2003qv}, 
or fundamental objects in string theory~\cite{Polchinski:2004ia}.
They might have been copiously produced in the early Universe and, 
if so, some of them are still present in the current Universe.
Detection of the cosmic string has profound implications on high energy physics
not probed by the terrestrial experiments. 
For instance, in the context of the field theory, 
measurement of the line density of the string directly enables us to 
estimate the energy scale of the symmetry breaking.
Various cosmological searches have been carried out to detect the cosmic strings
both directly and indirectly.
A straight cosmic string induces the gravitational lens effect and produce 
double images with exactly the same shapes.
Cosmic strings also source the temperature anisotropies of the cosmic microwave background radiation (CMBR).
Gravitational waves emitted from the cosmic strings modulate the pulsar timing and this fact 
has been used to look for the strings by the pulsar timing observations.
So far, none of the observations has succeeded in detecting the cosmic string and only the 
upper bound on the string line density has been imposed.
The latest CMBR observations by Planck satellite places the upper bound on the line density $U$ 
as $U < U_{\rm max} \equiv 1.8 \times 10^{20}~{\rm kg/m}$~\cite{Ade:2013xla}, 
assuming the Nambu-Goto string.
Observations of the pulsar timing put the stronger upper bound as $U<5.4 \times 10^{18}~{\rm kg/m}$,
although some assumptions about the string loop formation and gravitational wave emission are
made to derive this bound~\cite{vanHaasteren:2011ni}.
There are some inflationary models in which the cosmic strings with line density
as large as $U_{\rm max}$ do not conflict with the pulsar timing observations~\cite{Kamada:2012ag}.
Hereafter, we will assume that $U_{\rm max}$ is the maximally allowed 
value of the string line density.
Improvement of the sensitivity of any measurement in future has 
potential to detect the cosmic string. 

In this paper, instead of studying the cosmological implications of the cosmic string,
we consider passage of the cosmic string through the Earth and investigate the resultant distortion
of the Earth in order to determine what kind of signatures are expected to detect the string. 
Similar investigations are performed in~\cite{Khriplovich:2007ci,Khriplovich:2008er},
where the passage of the primordial black hole through the Earth is assumed,
and in~\cite{Pospelov:2012mt}, where the passage of the domain wall is considered.
As we will see, the cosmic string induces devastating global earthquake for the line density 
comparable to $U_{\rm max}$.  
Even for the much smaller line density, the distinctive and significant earthquake happens,
which may be detected by instruments such as seismometer and global positioning system (GPS).
We will see that we can detect the passage of the cosmic string of a line density down to ten orders of magnitude smaller than $U_{\rm max}$. This analysis can be straightforwardly extended to the main sequence stars.
We also discuss the string passage between two bodies such as the Earth-Moon system and gravitational interferometer.

\begin{figure}[t]  
\centering  
\includegraphics[width=77mm]{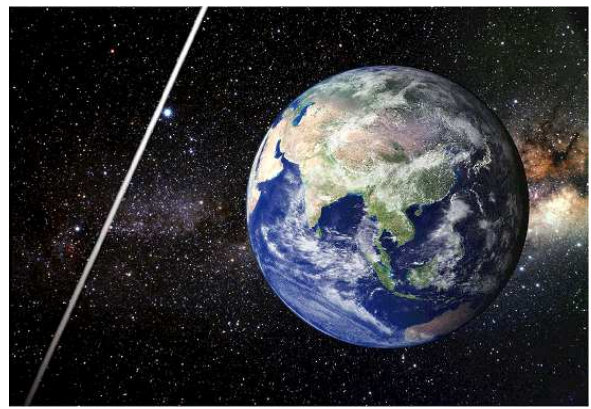}
\caption{
Passage of the cosmic string through the Earth. 
It causes the characteristic global earthquake from which we can detect the cosmic string 
with much smaller line density than the current cosmological constraint.
}  
\label{fig:es}  
\end{figure}  

\section{Passage of the cosmic string}
The spacetime around the static and straight Nambu-Goto string is locally Minkowski spacetime
with a deficit angle given by $\Delta =8\pi GU/c^2$ around the string~\cite{vilenkin-shellard}. 
Because of the deficit angle, 
when the string passes with speed $V$ between the two mass points that are at rest,
the mass points start to get closer each other by acquiring a constant relative velocity given by
$V\Delta=8\pi GUV/c^2$.
Interestingly, despite the acquisition of the non-zero relative velocity,
each mass point does not feel any force throughout the passage of the string, thus
being insensitive to the existence of the string unless some interaction with another
mass point is in operation.

Having this property in mind, let us consider a passage of the string through the Earth.
In order to obtain a qualitative understanding of what happens to the Earth after the string passage, 
we will make the following two simplifications.
First, we approximate the shape of the Earth by a cube with its size equal to the diameter of the Earth
and treat the Earth as uniform solid having no characteristic internal structure.
The internal structures such as the crust, the mantle and the core and the inhomogeneous
density and chemical compositions will induce various new effects that are not captured 
by our simple analysis.
In order to study those effects, massive numerical simulations are required,
which is far beyond the scope of this paper.   
These crude approximations allow us to do all the calculations analytically.
Secondly, we assume that the string slices the Earth in half 
in a way such that the resultant section is exactly parallel to the side of the cube.
Since the string motion is semi-relativistic $V ={\cal O}(0.1~c)$ which is much bigger than
the sound velocity $c_s$ of the material inside the Earth, we can safely regard the passage of
the string as an instant event ({\it i.e.}, we can say it occurs at $t=0$).

The most convenient Cartesian coordinate for our purpose is to put the center of the cube at the origin
and let sides of the cube be $x=\pm R,~y=\pm R$ and $z=\pm R$.
We can further set the section sliced by the string is the one specified by $x=0$.
Thus, the direction of the string motion lies on the $y-z$ plane and the cube contracts 
along the $x$-axis toward $x=0$ plane with initial speed given by 
\begin{equation}
v=\frac{4\pi GUV}{c^2}=28~{\rm m/s} \left( \frac{U}{U_{\rm max}} \right) 
\left( \frac{V}{0.1 c} \right). \label{st-ini-vel}
\end{equation}
The displacement vector ${\vec u}=u{\vec e_x}$ of the matter is along the $x$-axis,
which means that the longitudinal mode is excited.
The wave equation for $u(t,x)$ is given by
\begin{equation}
\frac{\partial^2 u}{\partial t^2}-c_s^2\frac{\partial^2 u}{\partial x^2}=v \delta(t)( \theta (-x)-\theta (x)), \label{wave-eq}
\end{equation}
where $\delta(t)$ is the Dirac delta function and $\theta(x)$ is the Heaviside step function.
The initial condition just before $t=0$ is $u=\partial_t u=0$.
The solution of Eq.~(\ref{wave-eq}) is given by
\begin{equation}
u(t,x)=v \int dx' ~G(t,0;x,x') (\theta (-x')-\theta (x')), \label{sol-1}
\end{equation}
where $G(t,t';x,x')$ is the retarded Green's function given by
\begin{equation}
G(t,t';x,x')=\frac{1}{2\pi c_s} \int dk~\frac{\sin c_s k(t-t')}{k} e^{-ik (x-x')}. \label{green}
\end{equation}
Substituting Eq.~(\ref{green}) into (\ref{sol-1}), we find that
\begin{align*}
u(t,x) & = \begin{cases}
                 vt, & x<-c_st\\
                 -\frac{v}{c_s}x, & -c_s t \le x \le c_s t \\
                 -vt.  & x>c_s t
           \end{cases}
\end{align*}
\begin{figure}[t]  
\centering  
\includegraphics[width=77mm]{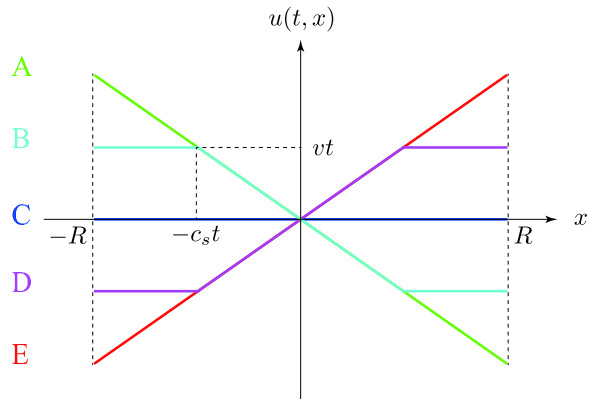}
\caption{
The displacement $u(t,x)$ of the matter. The oscillation proceeds through C $\to$ B $\to$ A $\to$ B' $\to$ C' $\to$ D' $\to$ E $\to$ D $\to$ C. After the passage of the cosmic string, two shocks propagate with the sound speed $c_s$. The matter inside the shocks is at rest and the matter outside the shocks is moving.
For the cases C, B and D, matter outside the shocks is shrinking, while it is expanding for the cases C', B' and D'. For the cases A and E where
the shocks reach the boundaries at $|x|=R$ and
the Earth becomes the minimum/maximum size $R(1\pm v/c_s)$, 
all the matter is at rest and the surface of the Earth receives impulsive force outward/inward, respectively.
}  
\label{fig:u}  
\end{figure}  
\begin{figure*}[t]  
\centering  
\includegraphics[width=140mm]{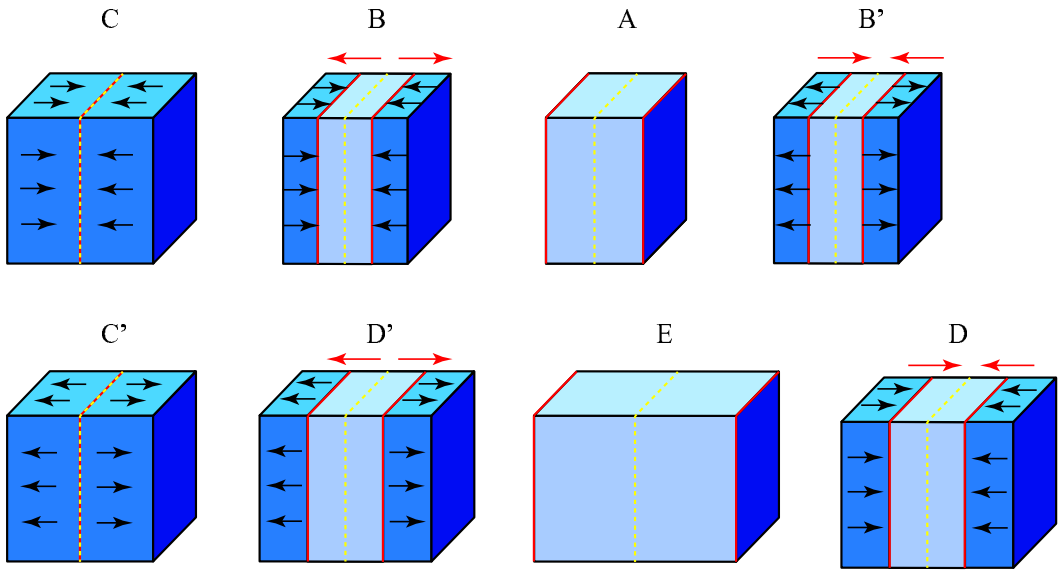}
\caption{
Schematic representation of Fig.~\ref{fig:u}.
The Earth is modeled as a cube with uniform density.
}  
\label{fig:cube}  
\end{figure*}  
Figure~\ref{fig:u} and \ref{fig:cube} represent time evolution of the Earth after the passage of the cosmic string. 
At $t=0$, $u(0,x)=0$ for any $|x|<R$ and the matter has the initial speed $v$ toward $x=0$. This is shown as the case C in Fig.~\ref{fig:u} and \ref{fig:cube}. 
The case B shows $u(t,x)$ in the time interval $0<t<R/c_s$.
The solution shows that there are two shocks located at $x=\pm c_s t$
and propagating at the sound speed $c_s$.
Notice that the matter outside the shocks ({\it i.e.}, $|x| > c_s t$) 
is still contracting with the initial speed $v$ toward $x=0$ and
the matter inside the shocks ({\it i.e.}, $|x| < c_s t$) is at rest.
Therefore, at the moment when the shocks reach the surface $x=\pm R$,
namely the case A,
the cube has shrunk along the $x$-axis by $Rv/c_s$ compared to the original size and
all the matter is at rest, which implies that the initial kinetic energy has been 
fully converted into the gradient energy.
After this time, the shocks turn back and move toward $x=0$ with the sound speed $c_s$. 
Since the surface of the cube is free end, 
it receives outward impulsive force when the shocks are at the surface.  
In terms of the velocity, the surface acquires outward $2v$ instantly, 
which results in the outward motion of the surface with a net speed $v$. 
Until the shocks reach again the center, matter outside the shocks expands with the speed $v$.
Thus, the displacement $u(t,x)$ is the same as that in the case B, 
but in this case the speed of the matter is opposite. 
We distinguish them by calling the latter case B', 
where the prime denotes that matter outside the shocks is expanding.
When the shocks arrive at the center, the cube comes back to the original size but all the matter
in the cube is moving outward with the speed $v$ (the case C').
After this time, the shocks move again outward and the region passed by the shocks comes to rest
with respect to the original center of mass frame, which is the case D'.
When the shocks reach the surface again, 
the whole cube comes to rest and its size along the $x$-axis is bigger than the original one
by $Rv/c_s$.
At this moment (case E), the radius of the Earth has reached the maximum.
The shocks are then reflected by the surfaces and move back to $x=0$.
The surface receives impulsive force toward the center upon the reflection of the shocks
and starts to shrink with speed $v$.
Therefore, we denote this case as D.
In summary, the shocks travel back and forth and the surface of the cube receives impulsive force each
time the shocks turn back at the surface.

The energy $E$ injected to the cube by the passage of the string is equal to the initial kinetic
energy of matter, $\frac{1}{2}\rho {(2R)}^3 v^2$, which is given by
\begin{equation}
E=4.4 \times 10^{27}~{\rm J}~ {\left( \frac{U}{U_{\rm max}} \right)}^2 
{\left( \frac{V}{0.1 c} \right)}^2 {\left( \frac{R}{R_\oplus} \right)}^3,
\end{equation}
where $R_\oplus = 6400~{\rm km}$ is the radius of the Earth.
This energy for the case of $U=U_{\rm max}$ amounts to kinetic energy of
meteor of $170~{\rm km}$ radius moving with the escape velocity $11.2~{\rm km/s}$.

The simple analysis above shows that string passage through the Earth causes 
the global oscillations of the Earth whose relative amplitude is the order of $v/c_s$.
We find that each oscillation is pulse-like, namely, the surface acquires instant 
acceleration at each turn and moves with a constant velocity otherwise.
In reality, the pulse-like oscillations become smooth due to the damping of 
high frequency modes and we end up with sinusoidal oscillations having periods of
${\cal O}(10^2)-{\cal O}(10^3)~{\rm s}$~\cite{ERI}. 
What is typically expected for the seismometer located on the surface is that it perceives
acceleration $a$ due to the sinusoidal oscillations estimated as $\frac{4\pi^2R}{T^2} \frac{v}{c_s}$, namely,
\begin{equation}
a=2~{\rm m/s^2} \left( \frac{U}{U_{\rm max}} \right) 
\left( \frac{V}{0.1 c} \right) {\left( \frac{T}{10^3~{\rm s}} \right)}^{-2} {\left( \frac{c_s}{3~{\rm km/s}} \right)}^{-1}.
\label{acce}
\end{equation}
Obviously, $2~{\rm m/s^2}$ is large enough acceleration to be perceived by human beings
without any sophisticated instrument and is expected to cause various dramatic
effects on the Earth.
The acceleration becomes smaller for smaller $U$.
The minimum acceleration that can be measured by the accelerograph is 
the order of $10^{-11}~{\rm m/s^2}$~\cite{ERI}.
Therefore, interestingly, by using the accelerograph we can detect the cosmic string with a tiny line density as small as $10^9~{\rm kg/m}$, which can never be detected by any cosmological observation.
We can even further reduce the minimum of the detectable $U$ by performing the matched filtering
analysis for the data and the template of the Earthquake caused by the string~\cite{ERI}.
In order to clarify how much the minimum of the detectable $U$ can be reduced, 
we need to do the simulations of the Earthquake with the initial condition
set by the string passage to prepare the theoretical template of the Earthquake
induced by the cosmic string.
However, this is a heavy task and beyond the scope of this paper.

GPS is another useful method to detect the oscillations of the Earth.
Oscillation amplitude induced by the string is
\begin{equation}
\frac{Rv}{c_s}=60~{\rm km}~\left( \frac{U}{U_{\rm max}} \right) 
\left( \frac{V}{0.1 c} \right). \label{gps-amp}
\end{equation}
The accuracy of the GPS is the order of $1~{\rm cm}$~\cite{ERI}.
Thus, the minimum $U$ detected by the GPS is about $10^{13}~{\rm kg/m}$
which is not as small as the one set by the accelerograph but is still much
lower than the cosmologically relevant value.

We can crudely estimate the rate at which the cosmic string passes through the Earth 
in a following way.
Let us suppose that there are $p$ long cosmic strings in the present Universe.
The value of $p$ is highly model dependent. 
So let us be open-minded and only require a trivial bound that the string energy 
density does not exceed the total energy density of the Universe.
This requirement yields an upper bound on $p$ as
\begin{equation}
p <\frac{c^2}{GU} =10^7~{\left(\frac{U}{U_{\rm max}} \right)}^{-1}. \label{pmax}
\end{equation}
Strings move over the cosmological distance ($\sim 10^{26}~{\rm m}$) 
on the order of the Hubble time $\sim 100~{\rm Gyr}$.
Thus, the probability $P$ that the string passage occurs within the Hubble time is estimated as
\begin{equation}
P \sim p \times \frac{R_\oplus }{10^{26}~{\rm m}} <
10^{-2} ~{\left( \frac{U}{10^{-10}U_{\rm max}} \right)}^{-1},
\end{equation}
where we have used Eq.~(\ref{pmax}) to derive the last inequality.
As we will see later, the cosmic string with line density as small as
$U=3\times 10^{-18}U_{\rm max}$ may be detected by using space interferometers.
In such a case, the event rate is further boosted.

So far, we have focused on the string passage through the Earth and studied
the resultant global oscillations.
The same analysis can be applied to the string passage through the main sequence stars.
After the string passes the star, the star starts to oscillate with the amplitude
$\sim R_s v/c_s$, where $R_s$ is the star radius and $c_s$ is the sound speed
of the gas inside the star.
The period of the oscillations is basically given by the free fall time 
\begin{equation}
T_s \sim \frac{1}{\sqrt{G\rho}}=3\times 10^3~{\rm s}~ 
{\left( \frac{R}{R_\odot}\right)}^{3/2} {\left( \frac{M}{M_\odot} \right)}^{-1/2},
\end{equation}
where $R_\odot$ and $M_\odot$ are the radius and the mass of the sun,
respectively.
The time-varying radius of the star results in the time-varying luminosity
due to the change of temperature roughly given by
\begin{equation}
\frac{\delta L_s}{L_s} \sim \frac{\delta R_s}{R_s} =10^{-4}~
\left( \frac{U}{U_{\rm max}} \right)
{\left( \frac{R}{R_\odot}\right)}^{3/2} {\left( \frac{M}{M_\odot} \right)}^{-1/2}.
\end{equation}
High-performance telescopes such as the Kepler spacecraft can measure $\delta L_s/L_s$
down to $10^{-5}$~\cite{Kepler}.
Therefore, if the string line density has a cosmologically interesting magnitude,
the effects of the string passage through the stars can be,
in principle, detected.

Finally, besides the direct passage through the Earth or stars, let us
consider the string passage between two  
bodies, {\it e.g.}, the Earth and the Moon.
First, we estimate the change of motion of two bodies with masses $m_1$ and $m_2$ 
that are orbiting each other by the gravitational force
after the cosmic string passes between the two bodies.
For simplicity, we assume that the initial orbit is a circle with 
radius $r_0$ and the string intersects perpendicularly 
with a line connecting the two bodies.
After the string passage, the two bodies start to approach each other
with a relative velocity given by Eq.~(\ref{st-ini-vel}).
As a result, the initial circular orbit becomes elliptic,
namely, the distance between the two bodies changes periodically. 
Treating this deviation as the first order perturbation,
a standard mechanical calculation gives the change of distance as
(assuming $m_1 \gg m_2$),
\begin{equation}
\frac{r_0^{3/2}v}{\sqrt{Gm_1} \pi^{3/4}}=4\times 10^3~{\rm km}
~\left( \frac{U}{U_{\rm max}} \right) {\left( \frac{m_1}{M_\oplus}\right)}^{-1/2}
{\left( \frac{r_0}{r_{\rm LD}} \right)}^{3/2},
\end{equation}
where $M_\oplus=6.0\times 10^{24}~{\rm kg}$ is the mass of the Earth and $r_{\rm LD}=3.8\times 10^5~{\rm km}$ is the lunar distance.
The Lunar Laser Ranging Experiment measures a distance between
the Earth and the Moon with an error of ${\cal O}({\rm cm})$~\cite{apollo}.
Therefore, this experiment can in principle detect the string passage between
the Earth and the Moon for the string line density down to $U \sim 10^{12}~{\rm kg/m}$.
Another example is that the cosmic string passes through space
interferometer such as DECIGO~\cite{Seto:2001qf} which aims to detect the gravitational waves.
DECIGO, whose arm length is about $10^3~{\rm km}$, is sensitive to the change of the arm length
down to $\sim 10^{-16}~{\rm m}$ per second.
Thus, using Eq.~(\ref{st-ini-vel}), we find that DECIGO may detect 
the passage of the cosmic string if its line density is larger than 
$3\times 10^{-18}U_{\rm max}$,
which is eight orders of magnitude stronger than that achieved by the accelerograph.

\section{Conclusion}
We studied the effects upon the Earth induced by the cosmic string when it passes through the Earth.
The passage of the cosmic string plays a role of providing initial conditions
for material such that two parts of the Earth separated by the plane spanned by the string trajectory
start to approach each other with a velocity proportional to the string line density.
Approximating the Earth as the uniform elastic cube, we found that the Earth then 
undergoes oscillations with average acceleration and amplitude given by 
Eq.~(\ref{acce}) and Eq.~(\ref{gps-amp}), respectively.
For the line density corresponding to the upper bound imposed by the CMBR observations,
these values are large enough to be perceived by human beings even
without using any sophisticated instrument.
On the other hand, 
even the cosmic string having the line density as small as $U =10^9~{\rm kg/m}$, 
which is ten orders of magnitude smaller than the cosmologically relevant value,
causes the oscillations of the Earth that may be measured by using the accelerograph.
Similar oscillations are also provoked when the cosmic string passes through the main
sequence star.
The oscillations of the star radius results in the periodic change of the luminosity.
The modulation of the luminosity becomes $\sim 10^{-4}$ for the string line density
corresponding to the upper bound from the CMBR observations, 
which may be detected by using the powerful contemporary telescopes.
The cosmic string may also trigger the periodic change of distance between 
two astronomical bodies if the string passes between them.
For the Earth-Moon system, the Lunar laser ranging experiment can detect the
string with line density as small as $U \sim 10^{12}~{\rm kg/m}$.
By using DECIGO, the minimum detectable value of $U$ can be as low as $10^2~{\rm kg/m}$.

\noindent {\bf Acknowledgments:} 
The authors thank to Teruyuki Kato and people in the Earthquake Research 
Institute at the University of Tokyo for letting us know many aspects of
the Earthquake.
We also thank to Toshikazu Shigeyama and Yousuke Itoh for useful comments.
This work is supported by Grant-in-Aid for Scientific Research on Innovative Areas
No.~25103505 (TS) from The Ministry of Education, Culture, Sports, Science and Technology (MEXT).

\end{document}